\RequirePackage{fix-cm}
\documentclass[smallextended]{svjour3}       
\usepackage[english]{babel}
\smartqed  
\usepackage{amsmath}
\usepackage{amssymb}
\usepackage{lipsum}
\usepackage{dsfont}
\usepackage{multirow}
\usepackage{afterpage}
\usepackage{amsfonts}
\usepackage{bm}%
\usepackage{babel}
\usepackage{graphicx}
\usepackage{color}
\usepackage{graphicx}
\begin{document}

\title{Asymmetry-induced nonclassical correlation 
}

 
\author{ R. Muthuganesan \and
       V. K. Chandrasekar 
}


\institute{Center for Nonlinear Science and Engineering,  School of Electrical and Electronics Engineering, SASTRA Deemed University, Tamil Nadu, India\at
              \email{rajendramuth@gmail.com}
}

\date{Received: date / Accepted: date}

\maketitle

\begin{abstract}
In  quantum resource theory (QRT), asymmetry recognized as a valid resource for the advantage of various quantum information processing. In this paper, we establish resource theory of asymmetry using quantum Fisher information (QFI). By defining the average Fisher information  as a measure of asymmetry, it is shown that the discrepancy of bipartite global and  local asymmetries naturally induces the nonclassical correlation between the subsystems.  This measure satisfies all the necessary axioms of a faithful measure of bipartite quantum correlation. As an illustration, we have studied the proposed measure for an arbitrary pure state and Bell diagonal state. 

\end{abstract}

\section{Introduction}
 
In the last two decades, quantum resource theory  has attracted much attention owing to its potential applications in quantum information processing  and considerable effort has been dedicated to the  development and implications of quantum resource theory \cite{Chithambar}. Any quantum system can exhibit many interesting properties such as entanglement \cite{Einstein1935,Schrodinger1935}, nonlocality and quantum coherence, which are consequence of superposition principle  and representing fundamental departures from classical physics \cite{Nielsen}. These peculiar properties are considered as an essential resources for quantum information processing tasks. Quantification of such features is a worthwhile task in the framework of resource theory. In recent years, many quantum resources in different context have been proposed such as  entanglement, quantum correlation (beyond entanglement) \cite{Knill,Datta2005,Datta2005PRL,Henderson,Ollivier,Gediscord,MIN,MID,UIN,muthu1,muthu2}, quantum coherence \cite{Baumgratz,coherence,Muthucoh}, asymmetry \cite{Dowling,Vaccaro} etc.

Among the above, asymmetry plays an important role in the development of quantum information theory and deep understanding of quantum systems. The quantum state may or may not be invariant with respect to the action of  the  group and the asymmetry is defined as the degree of symmetry breaking. In general, the symmetry breaking,  is a fundamental mechanism for the variety of matters, which is also useful in understanding of microscopic picture of fundamental interactions and emergence of macroscopic structure. Further, the significance of asymmetry can effectively identified in some significant  condensed matter phenomena, such as accidental degeneracy and spontaneous symmetry breaking. With the development of qualitative characterizations of asymmetry, recently,  the researchers paid wide attention on the quantification of asymmetry also. In the context various measures of asymmetry have been introduced and explored in detail, such as the entropic measure of asymmetry \cite{Vaccaro,Wiseman2003}, robustness of asymmetry \cite{Piani2016}, asymmetry weight \cite{BU}, measure of asymmetry based on commutators \cite{Fang,Dong}, Frobenius norm based measure of asymmetry \cite{Yao}. 
 
The quantum correlation and its monogamy relation are the  primary tools for the understanding of nature  nonclassical properties of a quantum system. In general, asymmetry and correlations are the fundamental quantum resources  and ubiquitous in quantum physics.   In this work, we investigate the quantification of quantumness from the viewpoint of resource theory of asymmetry. By defining asymmetry of the bipartite quantum state, we propose  the  difference between the bipartite global and  local asymmetries naturally induces the nonclassical correlation between the subsystems as a quantum correlation measure of the bipartite state. This measure satisfies basic requirements of bipartite  quantum correlation measure.

The structure of the paper as follows. In Sec. 2, we present the review of quantum Fisher information. In Sec. \ref{Asym}, we establish the resource theory of asymmetry in terms of QFI.  The bipartite quantum correlation measure defined and characterized  in Sec. \ref{Correl}. Finally, the conclusions are given in Sec. \ref{Concl}.

\section{Overview on QFI}
\label{QFI}

Quantum Fisher information (QFI) is the most important and promising tool in quantum information theory to enhance the precision and efficiency of quantum metrology protocols. Further, QFI is also useful in quantification of resources such as quantum coherence and quantum correlation. The dynamics and behaviors of QFI  in  the physical systems are  analyzed recently. In general, QFI quantifies the amount of information contained in state $\rho$ with respect to any physical observable $K$ \cite{Helstrom} and is reduced to variance of the observable in the state $\rho$. For any arbitrary quantum state $\rho_{\theta} = U \rho U^{\dagger}$ that depends on the parameter $\theta$, we can define the QFI as
\begin{align}
\mathcal{F}(\rho_{\theta})=\frac{1}{4}\text{Tr}[\rho_{\theta} L_{\theta}^2],  \nonumber
\end{align}
where the state  $\rho_{\theta}$ can be obtained from an initial probe state $\rho$ subjected to a unitary transformation $U = \mathrm{e}^{iK\theta}$. In order to extract the knowledge about $\theta$ from the parametric state $\rho_{\theta}$, we set  generalized quantum measurements, namely positive-operator-valued measures (POVMs), $M=\{ M_k|M_k\geq 0,\sum_kM_k=\mathds{1}\} $. Here, $L_{\theta}$ is symmetric logarithmic derivative defined as the solution of the equation
\begin{align}
\frac{d\rho_{\theta}}{d\theta}=\frac{1}{2}( \rho_{\theta} L_{\theta}+L_{\theta}\rho_{\theta} ). \nonumber
\end{align}
 Following the spectral decomposition $\rho=\sum_{i}p_i|\psi_i\rangle \langle \psi_i|$, with $p_i\geq 0$ and $\sum_{i=1}p_i=1$, the $ \mathcal{F}(\rho, K)$ is computed as \cite{Braunstein,Modi}
\begin{align}
\mathcal{F}(\rho, K)=\frac{1}{2}\sum_{i\neq j}\frac{(p_i-p_j)^2}{p_i+p_j}|\langle \psi_i|K |\psi_j\rangle|^2.
\label{QFI}
\end{align}
 
The quantum Fisher information has the following theoretic-information  properties: 
\begin{itemize}
  \item[$\mathcal{F}$1.] QFI is a non-negative i.e., $\mathcal{F}(\rho, K)\geq 0$.
  \item[$\mathcal{F}$2.] Quantum Fisher information reduces to variance $V(\rho, K)$ for pure states, namely, $\mathcal{F}(\rho, K) =V(\rho, K)$. In general, if $\rho$ is mixed, then 
  \begin{align}
 0  \leq  \mathcal{F}(\rho, K) \leq V(\rho, K) \nonumber
  \end{align}
  \item[$\mathcal{F}$3.] $\mathcal{F}(\rho, K)$ is unchanged under any unitary transformation $U$ commuting with the observable $K$ \cite{Toth}, i.e.,  
\begin{align}
\mathcal{F}(\rho, K)=\mathcal{F}(U\rho U^{\dagger}, K).\nonumber 
\end{align}
   \item[$\mathcal{F}$4.] $\mathcal{F}(\rho, K)$ is convex, which means that  if  several  different  quantum systems  are  mixed,  the  information  content  of  the  resulting system is not larger than the average information content of the component systems \cite{Toth}. Mathematically, 
\begin{align}
  \mathcal{F}\left(\sum_n\lambda_n\rho_n, K\right)\leq\sum_n\lambda_n \mathcal{F}(\rho_n, K) \nonumber
\end{align}
where $\sum_n\lambda_n=1$, $\lambda_n\geq 0$ and $\rho_n$ are quantum states.

\item[$\mathcal{F}$5.] $\mathcal{F}(\rho, K)$ is independent of choice of orthonormal basis $\rho^{ab}=\rho^a\otimes\rho^a$.

\item[$\mathcal{F}$6.] For bipartite state the Fisher information is superadditive \cite{Li}i.e., 
\begin{align}
\mathcal{F}(\rho^{ab}, K^a\otimes\mathds{1}^b+\mathds{1}^a \otimes K^b)\leq\mathcal{F}(\rho^a,K^a)+\mathcal{F}(\rho^b,K^b). \nonumber
\end{align}
The equality holds only for product state.
\end{itemize}
\section{Resource theory of asymmetry}
\label{Asym}
In quantum mechanics, any resource theory consists of a set of free states and a set of free operations. The amount of resource contained in a state can be quantified using resource monotones. The free states are not possessing the resource under consideration  and can not be used for quantum information processing tasks. On the other hand, free operations are quantum operations that map a free state into another free state or unable to create resource from the free states. In order to develop the resource theory of asymmetry based on QFI, we recall the definition of free states and free operations with in the framework of resource theory.  

Let $D(\mathcal{H})$ be the convex set of density operators acting on the Hilbert space $\mathcal{H}$ and $G$ is a symmetry group  with associated unitary representation $\{U_g\} $. Then we define, 
\begin{align}
\mathcal{U}_g(\tau)=U_g\tau U_g^{\dagger}
\end{align}
and the state $\sigma \in D(\mathcal{H})$ is defined as a symmetric state with respect to $G$ if and only if
\begin{align}
\mathcal{U}_g(\sigma)=\sigma
\end{align}
for all $g\in G$. Then we define a set which constitutes the set of free states for the resource theory of asymmetry as 
\begin{align}
\mathcal{S}:=\{ \sigma\in   D(\mathcal{H}):  \mathcal{U}_g(\sigma)=\sigma\}.
\end{align}
Similarly, a set of covariant operations with respect to group $G$ considered as  free operation in the resource theory of asymmetry. Such an operation is defined by a superoperator $\mathcal{L}: D(\mathcal{H} ) \rightarrow  D(\mathcal{H} )$ such that
\begin{align}
\mathcal{L}(\mathcal{U}_g(\tau))=\mathcal{U}_g(\mathcal{L}(\tau)), \forall ~~g \in G.
\end{align}

With the above perspective, we define asymmetry of arbitrary quantum state based on the QFI on the system space $\mathcal{H}$ with respect to Lie group $G\subset S(U)$  with dimension $d$. Let $\{T_j: j=1,2,3,\cdots, \text{dim} \mathcal{L}_G\}  $ be any orthonormal basis of $\mathcal{L}_G$, which is corresponding Lie algebra spanned by the generators of $G$ and a subspace of the Hilbert space consisting of all observables (Hermitian operators) on $K$ with the Hilbert-Schmidt inner product $\langle A |B\rangle=\text{Tr}AB $.  Here $\text{Tr}$ denotes trace of an operator. The asymmetry of the state $\rho$ with respect to the group $G$ or the corresponding Lie algebra $\mathcal{L}_G$ is defined as 
\begin{align}
\mathcal{A}(\rho, \mathcal{L}_G)=\frac{1}{4}\sum_j\mathcal{F}(\rho, T_j),
\end{align}
where the numerical factor $\frac{1}{4}$ is a normalization factor inserted for our convenience. It is worth mentioning that $\mathcal{A}(\rho, \mathcal{L}_G)$ is independent of  choice of the orthonormal basis $\{ T_j\}$ of $\mathcal{L}_G$ and considered as a well-defined measure. The QFI based measure of asymmetry of the state $\rho$ with respect to the group $G$ has the following interesting properties: 
\begin{itemize}
\item[i.] $\mathcal{A}(\rho, \mathcal{L}_G)\geq 0$. The equality holds if and only if $[\rho, T_j ] = 0, \forall ~j $.

\item[ii.] For any unitary operator $U$, $\mathcal{A}(\rho, \mathcal{L}_G)$ is unitary invariant in the sense that
\begin{align}
\mathcal{A}(\rho, \mathcal{L}_G)=\mathcal{A}(U\rho U^{\dagger}, \mathcal{L}_G). \nonumber
\end{align}
\item[iii.] $\mathcal{A}(\rho, \mathcal{L}_G)$ is convex with respect to $\rho$ in the sense that 
\begin{align}
\mathcal{A}\left(\sum_i \lambda_i \rho_i, \mathcal{L}_G\right)\leq \sum_i  \lambda_i \mathcal{A}\left(\rho_i, \mathcal{L}_G\right). \nonumber
\end{align}
where $\sum_i\lambda_i=1$, $\lambda_i\geq 0$ and $\rho_i$ are quantum states on the system space $H$.
\end{itemize}
The above properties can be seen directly from the properties of Fisher information.

\section{Asymmetry-induced nonclassical correlations}
\label{Correl}
In order to quantify the quantum correlation contained in a bipartite quantum state, we first define the asymmetry for bipartite quantum state. Let us consider a bipartite quantum state $\rho^{ab}$ on the composite separable Hilbert space $\mathcal{H}^{ab}=\mathcal{H}^a\otimes \mathcal{H}^b$ shared by $a$ and $b$. Here  $\rho^a$ and  $\rho^b$ are the marginal states on the corresponding Hilbert spaces $\mathcal{H}^a$ and $\mathcal{H}^b$ respectively with dimension $\text{dim}\mathcal{H}^{a(b)}=d_{a(b)}$. Then we define the orthonormal basis in the composite Hilbert space as 
\begin{align}
  \{X_m \otimes \mathds{1}^b,\mathds{1}^a \otimes Y_n: m=1,2,3, \cdots, d^2_a;~~ n=1,2,3, \cdots d^2_b\}, \nonumber
\end{align}
where $\{X_m: m=1,2, \cdots, d_a^2 \}$ and  $\{Y_n: n=1,2, \cdots, d_b^2 \}$ are any orthonormal bases of the Lie algebras $\mathcal{L}^a$ of $U(H^a)$ and  $\mathcal{L}^b$ of $U(H^b)$ respectively. 

With these settings, one can define asymmetry of a bipartite state as average Fisher information with respect to each generator of the basis. The measure of asymmetry of $\rho^{ab}$ with respect to the group $U (H^a) \times U (H^b)$, or equivalently, the corresponding Lie algebra $\mathcal{L}^{ab}=\mathcal{L}^{a}\oplus \mathcal{L}^{b}$:
\begin{align}
  \mathcal{A}(\rho^{ab},\mathcal{L}^{ab})=\frac{1}{4}\sum_{m=1}^{d_a^2}\mathcal{F}(\rho^{ab},X_m \otimes \mathds{1}^b)+\frac{1}{4}\sum_{n=1}^{d_b^2}\mathcal{F}(\rho^{ab},\mathds{1}^a \otimes Y_n). \nonumber
\end{align}
 We can rewrite the asymmetry of bipartite state as  
\begin{align}
  \mathcal{A}(\rho^{ab},\mathcal{L}^{ab})=\sum_{m=1}^{d_a^2}\mathcal{A}(\rho^{ab},\mathcal{L}^{a})+\sum_{n=1}^{d_b^2}\mathcal{A}(\rho^{ab},\mathcal{L}^{b}).
  \label{asymmetry}
\end{align}
Similarly, we can define asymmetry of local state $\rho^a=\text{Tr}_b(\rho^{ab})$ with respect to the group $U (H^a)$, and the corresponding Lie algebra $\mathcal{L}^{a}$ as 
\begin{align}
  \mathcal{A}(\rho^{a},\mathcal{L}^{a})=\frac{1}{4}\sum_{m=1}^{d_a^2}\mathcal{F}(\rho^{a},X_m).
\end{align}
The asymmetry of the local state $\rho^b=\text{Tr}_a(\rho^{ab})$ with the corresponding Lie algebra $\mathcal{L}^{b}$ is
\begin{align}
  \mathcal{A}(\rho^{b},\mathcal{L}^{b})=\frac{1}{4}\sum_{n=1}^{d_b^2}\mathcal{F}(\rho^{b},Y_n).
\end{align}
The asymmetries of the local states  $\mathcal{A}(\rho^{a},\mathcal{L}^{a})$ and $\mathcal{A}(\rho^{b},\mathcal{L}^{b})$ are also independent of the choice of the orthonormal bases of $\mathcal{L}^{a}$ and $\mathcal{L}^{b}$ respectively.  The sum of local asymmetries quantifies the total asymmetry of local states. 

The measure of asymmetry of the global state $ \mathcal{A}(\rho^{ab},\mathcal{L}^{ab})$ with respect to the local unitary group has the following interesting properties: 

\begin{itemize}
\item[$\mathcal{A}$1.] $\mathcal{A}(\rho^{ab},\mathcal{L}^{ab})\geq 0$, equality holds if and only if $\rho^{ab}$ is a maximally mixed state i.e., $\rho^{ab}=\mathds{1}^{ab}/d_ad_b$, where $\mathds{1}^{ab}$ is the identity operator on the bipartite system space $H^a \otimes H^b$.

\item[$\mathcal{A}$2.] $\mathcal{A}(\rho^{ab},\mathcal{L}^{ab})$ is locally unitary invariant i.e., 
\begin{align}
\mathcal{A}\left(\rho^{ab},\mathcal{L}^{ab}\right)= \mathcal{A}((U\otimes V)\rho^{ab}(U\otimes V)^{\dagger},\mathcal{L}^{ab})   \nonumber
\end{align}

\item[$\mathcal{A}$3.]  $\mathcal{A}(\rho^{ab},\mathcal{L}^{ab})$ is convex with respect to $\rho^{ab}$ in the sense that 
\begin{align}
\mathcal{A}\left(\sum_j\lambda_j\rho^{ab}_j,\mathcal{L}^{ab}\right)\leq \sum_j \lambda_j\mathcal{A}(\rho^{ab}_j,\mathcal{L}^{ab})   \nonumber
\end{align}
\item[$\mathcal{A}$4.] $\mathcal{A}(\rho^{ab},\mathcal{L}^{ab})$ is superadditive in the sense that
\begin{align}
\mathcal{A}(\rho^{ab},\mathcal{L}^{ab})\geq \mathcal{A}(\rho^{a},\mathcal{L}^{a})+\mathcal{A}(\rho^{b},\mathcal{L}^{b}). \nonumber
\end{align}
The equality holds only for the product state. 
\item[$\mathcal{A}$5.] $\mathcal{A}(\rho^{ab},\mathcal{L}^{ab})$ is not increased under the completely positive and trace preserving (CPTP) map.

\item[$\mathcal{A}$6.] $\mathcal{A}(\rho^{ab},\mathcal{L}^{ab})$ is independent of   choice of orthonormal basis. 
\end{itemize}
Now we sketch the proof of the above properties. 

It is quite easy to understand the positivity of $\mathcal{A}(\rho^{ab},\mathcal{L}^{ab})$ from the positivity of Fisher information. From Eq. (\ref{asymmetry}), we can understand the fact that $\mathcal{A}(\rho^{ab},\mathcal{L}^{ab})=0$ leads $\mathcal{A}(\rho^{ab},\mathcal{L}^{a})=0$ and $\mathcal{A}(\rho^{ab},\mathcal{L}^{b})=0$. Using the result given in ref. [], we can observe that 
\begin{align}
\mathcal{A}(\rho^{ab},\mathcal{L}^{ab})- \mathcal{A}(\rho^{a},\mathcal{L}^{a})\geq 0.
\label{diffasy}
\end{align}
For product state $\rho^{ab}=\rho^{a}\otimes \rho^{b}$, the asymmetry of global state is equal to its marginal asymmetry i.e., $\mathcal{A}(\rho^{ab},\mathcal{L}^{ab})- \mathcal{A}(\rho^{a},\mathcal{L}^{a})$. Hence, the above equality holds for only product state. If $\mathcal{A}(\rho^{ab},\mathcal{L}^{ab})=0$, from the above equation, we find that $\mathcal{A}(\rho^{a},\mathcal{L}^{a})=0$ and this is the case only when $\rho^{a}=\mathds{1}/d_a$ is the maximally mixed state. Similarly, we conclude that $\rho^b = \mathds{1}^b/d_b$ is also the maximally mixed state. Combining these facts, it follows that  is $\rho^{ab}=\mathds{1}^{ab}/d_ad_b$ the maximally mixed state on $H^a\otimes H^b$ .

Properties $\mathcal{A}$2--$\mathcal{A}$4 follows straightforwardly from the properties of Fisher information $\mathcal{F}$3--$\mathcal{F}$5. To show the property $\mathcal{A}$5, we have to show that the quantum Fisher information do not increase under CPTP map. To show this, First we define a CPTP map on subsystem $b$ and it can be expressed as
\begin{align}
\mathcal{I}^a\otimes\mathcal{E}^{ab}(\rho^{ab})=\text{Tr}_c(\mathds{1}^a\otimes U^{bc}\rho^{ab}\otimes\rho^c(\mathds{1}^a\otimes U^{bc})^{\dagger})
\end{align}
where $\rho^c$ is a density matrix of an ancillary system $c$, and $U^{bc}$ is a unitary operation in state space of composite system $bc$.
\begin{eqnarray}
\mathcal{F}(\mathcal{I}^a\otimes\mathcal{E}^{ab}(\rho^{ab}, T^a\otimes\mathds{1}^b)=\mathcal{F}(\text{Tr}_c(\mathds{1}^a\otimes U^{bc}\rho^{ab}\otimes\rho^c(\mathds{1}^a\otimes U^{bc})^{\dagger},T^a\otimes\mathds{1}^b) \nonumber \\
\leq \mathcal{F}(\mathds{1}^a\otimes U^{bc}\rho^{ab}\otimes\rho^c(\mathds{1}^a\otimes U^{bc})^{\dagger},T^a\otimes\mathds{1}^b) ~~~~~ \nonumber \\
=\mathcal{F}(\rho^{ab}\otimes\rho^c, \mathds{1}^a\otimes U^{bc} (T^a\otimes\mathds{1}^b)\mathds{1}^a\otimes U^{bc})^{\dagger}) ~~~~\nonumber \\
=\mathcal{F}(\rho^{ab}\otimes\rho^c,T^a\otimes\mathds{1}^b) ~~~~~~~~~~~~~~~~~~~~~~~~~~~~~~~~\nonumber \\
=\mathcal{F}(\rho^{ab},T^a\otimes\mathds{1}^b). ~~~~~~~~~~~~~~~~~~~~~~~~~~~~~~~~~~~~~\nonumber
\end{eqnarray}
Here the inequality follows from Eq. (\ref{diffasy}). 

Next, we define bipartite quantum correlation measure in terms of asymmetry of bipartite state. In general, the asymmetry of $\rho^{ab}$ with respect to $\mathcal{L}^a$ is larger than the asymmetry of $\rho^a$ with respect to $\mathcal{L}^a$.  Similarly, the asymmetry of $\rho^{ab}$ with respect to $\mathcal{L}^b$ is larger than the asymmetry of $\rho^b$ with respect to $\mathcal{L}^b$. We define the quantum correlation as the discrepancy between the asymmetries of a bipartite state and its marginal states with respect to the same group that comes from the correlations contained in the bipartite state. Mathematically, it can be written
\begin{align}
\mathcal{Q}(\rho^{ab})=\mathcal{A}(\rho^{ab},\mathcal{L}^{ab})-\mathcal{A}(\rho^{a},\mathcal{L}^{a})-\mathcal{A}(\rho^{b},\mathcal{L}^{b})
\end{align}
as a measure of correlations contained in the state $\rho^{ab}$. Using the property of QFI  ($\mathcal{F}$4.), the correlation measure is redefined as 
\begin{align}
\mathcal{Q}(\rho^{ab})=\mathcal{A}(\rho^{ab},\mathcal{L}^{ab})-\mathcal{A}(\rho^{a}\otimes\rho^{b},\mathcal{L}^{ab}).
\end{align}
 Using Eq. (\ref{asymmetry}), it can be rewritten as 
\begin{eqnarray}
\mathcal{Q}(\rho^{ab})=\mathcal{A}(\rho^{ab},\mathcal{L}^{a})+\mathcal{A}(\rho^{ab},\mathcal{L}^{b})-\mathcal{A}(\rho^{a},\mathcal{L}^{a})-\mathcal{A}(\rho^{b},\mathcal{L}^{b}) \nonumber \\
=\mathcal{A}(\rho^{ab},\mathcal{L}^{a})-\mathcal{A}(\rho^{a},\mathcal{L}^{a})+\mathcal{A}(\rho^{ab},\mathcal{L}^{b})-\mathcal{A}(\rho^{b},\mathcal{L}^{b}) \nonumber \\
=\mathcal{Q}^a(\rho^{ab})+\mathcal{Q}^b(\rho^{ab})~~~~~~~~~~~~~~~~~~~~~~~~~~~~~~~~~~~~~~~
\end{eqnarray}
Here $\mathcal{Q}^{a(b)}(\rho^{ab})=\mathcal{A}(\rho^{ab},\mathcal{L}^{a})-\mathcal{A}(\rho^{a(b)},\mathcal{L}^{a(b)})$, which quantifies the discrepancy between 
the asymmetries between $\rho^{ab}$ and $\rho^{a(b)}$ with respect to the local unitary group on system $a(b)$. In general, a faithful measure of bipartite correlation should satisfy three necessary axioms such as (i) should be zero for tensor product states, and a positive for entangled states, (ii) should be invariant under local unitary transformations on each one of the two subsystems and (iii) should coincide with quantum entanglement for pure states. Here we demonstrate the some important properties of asymmetry based quantum correlation  measure $\mathcal{Q}(\rho^{ab})$: 
\begin{itemize}
\item[$\mathcal{Q}$1.] $\mathcal{Q}(\rho^{ab})\geq 0$. The equality hold for product state $\rho^{ab}=\rho^a\otimes \rho^b$.

\item[$\mathcal{Q}$2.] $\mathcal{Q}(\rho^{ab})$ is locally unitary invariant.

\item[$\mathcal{Q}$3.] $\mathcal{Q}(\rho^{ab})$  is convex with respect to $\rho^{ab}$ in the sense that 
\begin{align}
\mathcal{Q}\left(\sum_j\lambda_j\rho^{ab}_j,\mathcal{L}^{ab}\right)\leq \sum_j \lambda_j\mathcal{Q}(\rho^{ab}_j,\mathcal{L}^{ab}).   \nonumber
\end{align}
\item[$\mathcal{Q}$4.]  For any pure state $|\Psi\rangle$ with $d=d_a\leq d_b$, $\mathcal{Q}(|\Psi\rangle \langle \Psi|)\leq (d-1)/d$.  The maximum is achieved when the state is  maximally entangled.
\end{itemize}
The properties $\mathcal{Q}$1--$\mathcal{Q}$3 are straightforward to prove from the properties of asymmetry of  bipartite state. Hence, the asymmetry based quantum correlation is a valid quantifier of bipartite correlation. We shall provide  proof of the property $\mathcal{Q}$4 in the next section.  

Now, we generalize these formalisms to the resource theory of asymmetry for multipartite system and define a measure of multipartite quantum correlation measure using asymmetry. Let $\rho^{a_1, a_2, \cdots,a_n}$ be the multipartite states of the system space $\mathcal{H}^{a_1} \otimes \cdots \otimes \mathcal{H}^{a_n}$. The asymmetry of the state directly defined as 
\begin{align}
 \mathcal{A}(\rho^{a_1,\cdots,a_n},\mathcal{L}^{a_1, \cdots,a_n})=\sum_{i=1}^n\mathcal{A}(\rho^{a_1,\cdots,a_n},\mathcal{L}^{a_i}),
 \end{align}
where $\mathcal{L}^{a_1, \cdots,a_n}=\mathcal{L}^{a_1}\oplus   \cdots \oplus \mathcal{L}^{a_1}$ is the Lie algebra of the local unitary group $U(H^{a_1})\times \cdots \times U(H^{a_n})$ on the total space $H^{a_1} \cdots \otimes H^{a_n}$. The corresponding measure of correlations can be defined as
\begin{align}
\mathcal{Q}(\rho^{a_1,\cdots,a_n})=\mathcal{F}(\rho^{a_1,\cdots,a_n},\mathcal{L}^{a_1, \cdots,a_n})-\mathcal{F}(\rho^{a_1} \otimes \cdots \otimes\rho^{a_n}, \mathcal{L}^{a_1, \cdots,a_n}), \nonumber
 \end{align}
and in terms of asymmetry 
\begin{align}
\mathcal{Q}(\rho^{a_1,\cdots,a_n})=\sum_{i=1}^n\mathcal{A}(\rho^{a_1,\cdots,a_n},\mathcal{L}^{a_i})-\mathcal{A}(\rho^{a_i},\mathcal{L}^{a_i}).
\end{align}
The above measure quantifies the amount of correlations in multipartite quantum state $\rho^{a_1,\cdots,a_n}$ in terms of the differences between asymmetries of the global
and local states. 
\subsection{Examples}
In this section, we compute the asymmetry induced correlation measure for an arbitrary pure and Bell diagonal state.
 
\textit{Pure state:} The Schmidt decomposition of the pure state as 
\begin{align}
|\psi\rangle =\sum_{i=1}^d\sqrt{\lambda_i}|i\rangle_a|i\rangle_b, ~~~~~~ d=\text{min}\{ d_a, d_b\} ,
\end{align}
with reduced state $\rho^{a(b)}=Tr_{b(a)}(|\psi\rangle \langle \psi|)$ and $\boldsymbol\lambda_{|\psi\rangle}=(\lambda_1,\lambda_1, \cdots, \lambda_d)$ be the Schmidt coefficients, with $\lambda_i\geq 0$ and $\sum_i\lambda_i=1$. To compute $\mathcal{Q}(|\psi\rangle \langle \psi|)$, first we define the set of generators as 
\begin{eqnarray}
T_{ij}^{(1)}=\frac{1}{2}(|i\rangle \langle j|+|j\rangle \langle i|), ~~~~~~1\leq i <  j \leq d,~~~~~~~~~~~~~~~~~~~~~~~~~~ \nonumber \\ 
T_{ij}^{(2)}=\frac{1}{2\mathrm{i}}(|i\rangle \langle j|-|j\rangle \langle i|), ~~~~~~1\leq i <  j \leq d, ~~~~~~~~~~~~~~~~~~~~~~~~~~ \nonumber \\ 
T_{ij}^{(3)}=\frac{1}{2\sqrt{i(i-1)}}(|1\rangle \langle 1|+|2\rangle \langle 2|+\cdots+|i-1\rangle \langle i-1|+(1-i)|i\rangle \langle i|), ~~~~~~2\leq i \leq d. \nonumber
\end{eqnarray}
Then, we compute 
\begin{align}
\mathcal{F}(|\psi\rangle \langle \psi|)=\frac{1}{2}\left(d-\sum_i^d\lambda^2_i\right), ~~~~~~~\mathcal{F}(\rho^a)=\frac{1}{4}\left(2d-\sum_{i,j}\frac{4\lambda_i\lambda_j}{\lambda_i+\lambda_j}\right), \nonumber
\end{align}
and 
\begin{align}
\mathcal{Q}^a(|\psi\rangle \langle \psi|)=\frac{1}{2}\sum_{i\neq j}\left(\frac{2\lambda_i\lambda_j}{\lambda_i+\lambda_j}+\lambda_i\lambda_j\right). \nonumber
\end{align}
Considering the same set of generators for the subsystem $b$, we have 
\begin{align}
\mathcal{Q}^b(|\psi\rangle \langle \psi|)=\frac{1}{2}\sum_{i\neq j}\left(\frac{2\lambda_i\lambda_j}{\lambda_i+\lambda_j}+\lambda_i\lambda_j\right). \nonumber
\end{align}
Then 
\begin{align}
\mathcal{Q}(|\psi\rangle \langle \psi|)=\sum_{i\neq j}\left(\frac{2\lambda_i\lambda_j}{\lambda_i+\lambda_j}+\lambda_i\lambda_j\right). \nonumber
\end{align}
$\mathcal{Q}^a(|\psi\rangle \langle \psi|)$ is a Schur concave, implies that $\mathcal{Q}^a(|\psi\rangle \langle \psi|)\geq \mathcal{Q}^a(|\phi\rangle \langle \phi|)$ iff the Schmidt coefficient $\boldsymbol\lambda_{|\psi\rangle}$ is majorized by $\boldsymbol\lambda_{|\phi\rangle}$, which means that  $\sum_{l=1}^k\lambda^{\downarrow}_{|\psi\rangle}\leq \sum_{l=1}^k\lambda^{\downarrow}_{|\psi\rangle}$ with $k=1,2,\cdots,d$ and equality is required for $k = d$, where $\downarrow$ indicates rearranging elements in descending order. From Schur concavity, the bound of $\mathcal{Q}(|\psi\rangle \langle \psi|)$ is given as 
\begin{align}
0\leq \mathcal{Q}(|\psi\rangle \langle \psi|)\leq \frac{d-1}{d}.
\end{align}
Here the lower bound is reached by product states with $\boldsymbol\lambda^{\downarrow}_{|\psi\rangle}=(1,0,0, \cdots, 0)$ and the upper bound is reached by maximally
entangled states with $\boldsymbol\lambda^{\downarrow}_{|\psi\rangle}=(1/d,1/d, \cdots, 1/d)$. The concurrence of the pure state is $C(|\psi\rangle \langle \psi|)=2\sqrt{\lambda_1\lambda_2}$. The $\mathcal{Q}^a(|\psi\rangle \langle \psi|)$ is computed as 
\begin{align}
\mathcal{Q}^a(|\psi\rangle \langle \psi|)=3\lambda_1\lambda_2,
\end{align}
and the correlation is connected with the concurrence in the following way 
\begin{align}
\mathcal{Q}(|\psi\rangle \langle \psi|)=\frac{3}{2}C^2(|\psi\rangle \langle \psi|).
\end{align}

\textit{Bell diagonal state:} Next we study the asymmetry of the Bell diagonal state whose marginal states are maximally mixed. In Bloch representation of the state can be expressed as
\begin{align}
\rho=\frac{\mathds{1}}{4}+\sum_ic_i \sigma_i \otimes \sigma_i.
\end{align}
The asymmetries of the local states are $\mathcal{A}(\rho^a, \mathcal{L}^a)=\mathcal{A}(\rho^b, \mathcal{L}^b)=0$. The Fisher information of the global state is 
\begin{align}
\mathcal{F}(\rho, T)=3-4\sum_{i>j}\frac{\beta_i\beta_j}{\beta_i+\beta_j},
\end{align}
where $\beta_i$ are the eigenvalues of Bell diagonal state with 
\begin{align}
\beta_1=1/4-c_1+c_2+c_3, ~~~~~~~~~~~ \beta_2=1/4+c_1-c_2+c_3,   \nonumber \\
\beta_3=1/4+c_1+c_2-c_3, ~~~~~~~~~~~ \beta_4=1/4-c_1-c_2-c_3.  \nonumber
\end{align}
Then, 
\begin{align}
\mathcal{Q}^a(\rho)=\mathcal{Q}^b(\rho)=\frac{1}{4}\left(3-4\sum_{i>j}\frac{\beta_i\beta_j}{\beta_i+\beta_j}\right),  \nonumber 
\end{align}
and the quantum correlation is 
\begin{align}
\mathcal{Q}(\rho)=\frac{1}{2}\left(3-4\sum_{i>j}\frac{\beta_i\beta_j}{\beta_i+\beta_j}\right).
\end{align}

\section{Conclusions}
\label{Concl}

In conclusion, we have defined the asymmetry of state as average of quantum Fisher information with respect to Lie groups and Lie algebras.  As a tool on the way, we have shown that the discrepancy between the asymmetry of bipartite global state and asymmetries of local states quantifies the quantum correlation contained in a bipartite quantum state. Further, this quantity satisfies the axioms of a valid measure of bipartite quantum correlation and is nonincreasing  under the action of general CPTP quantum channels on one of the subsystems. Hence, the proposed quantity is a good measure of bipartite quantum correlation. 
 
Since the asymmetry-based quantum correlation measure does not have an optimization procedure,  is a computable measure irrespective of dimension of the Hilbert space of the subsystems. We have established a simple relation between the concurrence and the correlation measure. Moreover, we have  generalized the measure to multipartite settings. Finally, the significance of this measure was illustrated with some examples.

The quantum Fisher information is also closely connected with parameter estimation and useful in quantum metrology. The proposed correlation measure has more insight into metrology and quantum information theory.

\begin{acknowledgements}

This work has been financially supported by the Council of Scientific and Industrial Research (CSIR), Government of India for the financial
support under Grant No. 03(1444)/18/EMR-II.
\end{acknowledgements}



\end{document}